\documentstyle{article}
 \def\test#1{}     

\def\beq{\begin{equation}}
\def\eeq{\end{equation}}

\def\beqx{\begin{displaymath}}
\def\eeqx{\end{displaymath}}

\def\beql{\arraycolsep .5mm \begin{eqnarray}}
\def\eeql{\end{eqnarray}}
\def\zeile{\nonumber \\[2mm] }   

\def\labelx#1{\label{#1}\test
                    {\raisebox{-4ex}{\makebox[0pt][r]{\tt #1}}}}

\def\rf#1{(\ref{#1})}

\newcommand{\Section}[1]{\section[#1]{#1}
                         \setcounter{equation}{0}
   \renewcommand{\theequation}{\arabic{section}.\arabic{equation}}}

\newcommand{\Appendix}[2]{\section*{Appendix #1: #2}
                          \addcontentsline{toc}{section}{#1: #2}
                          \setcounter{equation}{0}
   \renewcommand{\theequation}{#1.\arabic{equation}}}

\def\wurzelhalb{{\textstyle  {1 \over \sqrt 2}}}
\def\wurzelzwei{{\textstyle  \sqrt 2}}
\def\halb{{\textstyle {1\over 2}}}
\def\ft#1#2{{\textstyle{{#1}\over{#2}}}}
\def\a{\alpha}
\def\b{\beta}
\def\g{\gamma}\def\G{\Gamma}
\def\d{\delta}

\def\e{\epsilon}

\def\t{\theta}
\def\l{\lambda}\def\L{\Lambda}

\def\n{\nu}
\def\p{\psi}
\def\r{\rho}

\def\o{\omega}\def\O{\Omega}

\def\ve{\varphi}
\def\pa{\partial}
\def\rD{{\rm D}}

\def\Adot{{\dot A}}
\def\Bdot{{\dot B}}

\def\thp#1{{\theta_{\dot +}^{#1}}}
\def\thm#1{{\theta_{\dot -}^{#1}}}
\def\smpm#1{ {\psi_{\dot +}^{{#1}{\dot -}}}}
\def\smmp#1{ {\psi_{\dot -}^{{#1}{\dot +}}}}
\def\pzp#1{{\psi_{2\dot +}^{#1}}}
\def\pzm#1{{\psi_{2\dot -}^{#1}}}
\def\chp#1{{\chi_{\dot +}^{#1}}}
\def\chm#1{{\chi_{\dot -}^{#1}}}

\def\ep#1{{\epsilon^{{\dot +} {#1}} }}
\def\em#1{{\epsilon^{{\dot -} {#1}} }}

\def\dop{{\dot +}}
\def\dom{{\dot -}}
\def\dxp{dx^{\dot +}}
\def\dxm{dx^{\dot -}}
\def\Dxp{{\cal D}x^{\dot +}}
\def\Dxm{{\cal D}x^{\dot -}}
\def\Dp{{\cal D}_{\dot +}}
\def\Dm{{\cal D}_{\dot -}}
\def\rDp{{\rm D}_{\dot +}}
\def\rDm{{\rm D}_{\dot -}}
\def\mpm{\mu_{\dot +}^{\;\; \dot -}}
\def\mmp{\mu_{\dot -}^{\;\; \dot +}}
\def\MM{1- \mu_{\dot +}^{\;\; \dot -} \mu_{\dot -}^{\;\; \dot +} }
\def\epp{e_{\dot +}^{\;\; +}}
\def\emm{e_{\dot -}^{\;\; -}}
\def\tepp{{\tilde e}_{\dot +}^{\;\; +}}
\def\temm{{\tilde e}_{\dot -}^{\;\; -}}
\def\iepp{e_+^{\;\; \dot +}}
\def\iemm{e_-^{\;\; \dot -}}
\def\itepp{{\tilde e}_+^{\;\; \dot +}}
\def\itemm{{\tilde e}_-^{\;\; \dot -}}
\def\opmp{ \tilde \omega_{\dot +}^{\;\; -+} }
\def\oppm{ \tilde \omega_{\dot +}^{\;\; +-} }
\def\ompm{ \tilde \omega_{\dot -}^{\;\; +-} }
\def\ommp{ \tilde \omega_{\dot -}^{\;\; -+} }
\def\Gpmm{ \tilde \Gamma_{\dot + \dot -}^{\;\;\;\;\; \dot -}}
\def\Gmpp{ \tilde \Gamma_{\dot - \dot +}^{\;\;\;\;\; \dot +}}
\def\Gppp{ \tilde \Gamma_{\dot + \dot +}^{\;\;\;\;\; \dot +}}
\def\Gmmm{ \tilde \Gamma_{\dot - \dot -}^{\;\;\;\;\; \dot -}}
\def\Gppm{ \tilde \Gamma_{\dot + \dot +}^{\;\;\;\;\; \dot -}}
\def\Gmmp{ \tilde \Gamma_{\dot - \dot -}^{\;\;\;\;\; \dot +}}
\def\te{{{\tilde e}}}

\def\GI#1#2{ \Gamma^I_{ {#1} {#2}} }
\def\GIJ#1#2{ \Gamma^{IJ}_{{#1} {#2}} }

\def\euo#1#2{  e_{#1}^{\;\; {#2}} }

\def\V    {  {\cal V} }
\def\Vinv {  \V^{-1}  }
\def\Vh    { \widehat {\cal V} }

\def\VdV_#1{  \Vinv \partial_{#1}  \V }
\def\VDV_#1{  \Vinv {cal D}_{#1}  \V }
\def\VdVh_#1{ \Vh^{-1} \partial_{#1}  \Vh }
\def\VDVh_#1{ \Vh^{-1}  {cal D}_{#1}  \Vh }
\def\cA{{\cal A}}

\def\cP{{\cal P}}
\def\cQ{{\cal Q}}
\def\cM{{\cal M}}
\def\cR{{\cal R}}
\def\cT{{\cal T}}
\def\cJ{{\cal J}}
\def\cS{{\cal S}}

\def\c#1{{\cal #1}}

\title{\sc New Linear Systems for 2d Poincar\'e Supergravities}
\author{\sl H. Nicolai \\
         II. Institut f\"ur Theoretische Physik  \\
         Universit\"at Hamburg  \\
         22761 Hamburg, F.R.G.}

\begin{document}
\maketitle
\begin{abstract}
A new linear system is constructed for Poincar\'e
supergravities in two dimensions. In contrast to previous
results, which were based on the conformal gauge,
this linear system involves the topological world sheet
degrees of freedom (the Beltrami and super-Beltrami differentials).
The associated spectral parameter likewise depends on these
and is itself subject to a pair of differential
equations, whose integrability condition yields one of the
equations of motion. These results suggest the existence of
an extension of the Geroch group mixing propagating and
topological degrees of freedom on the world sheet.
We also develop a chiral tensor formalism for arbitrary
Beltrami differentials, in which the factorization
of $2d$ diffeomorphisms is always manifest.
\end{abstract}
\vfill

\newpage

\Section{Introduction}

The purpose of this paper is to generalize the linear systems
(or Lax pairs) that were derived already some time ago for the
dimensionally reduced field equations of Einstein Yang-Mills
theories \cite{Maison, Hoens, BM} and
their locally supersymmetric extensions \cite{NW} (for
a recent review, see \cite{Schlad}). These reductions
correspond to solutions of the field equations, which depend on two
coordinates only and thus possess at least two commuting Killing
vectors. The models obtained in this way closely resemble flat space
integrable non-linear sigma models in two dimensions, and indeed the
associated linear systems constructed so far are almost identical (see
\cite{Pohlmeyer, EF} for the flat space models). The present work
differs from earlier treatments, which were all based on the
(super)conformal gauge, in that it allows for non-trivial
topologies of the two dimensional world sheets by taking into account
the topological degrees of freedom of the world sheet, i.e. its
moduli and supermoduli. These constitute extra physical
(but non-propagating) degrees of freedom not present
in the corresponding flat space integrable sigma models,
and affect the dynamics in a non-trivial fashion.
In particular, there is a ``back reaction" of the matter
fields on the topological degrees of freedom, in contrast to conformal
field theories, where the moduli determining the background can be
freely chosen. The spectral parameter $t$ entering the linear system
is now not only a function of the ``dilaton" field as in
\cite{BM, NW}, but also depends on the moduli and super-moduli
of the world sheet. It is subject to a pair of differential
equations, whose integrability condition yields one of the
equations of motion obtained by dimensional reduction of
Einstein's equations.

Despite the possible relevance of these results for the
construction of new solutions of Einstein's equations, such as the
wormhole type solutions proposed in \cite{KM}, a far more
important concern is the search for new symmetries
generalizing the Geroch group \cite{Geroch} and the ``hidden
symmetries" of dimensionally reduced supergravities \cite{CJ, Julia}.
To a large extent, the present investigation is motivated by
\cite{Julia, Julia2}, where the connection between $2d$
supergravities and infinite dimensional Lie algebras of Kac Moody
type was pioneered, and where it was shown that the
Geroch group in infinitesimal form is nothing but the affine
Kac Moody algebra $A_1^{(1)}$, i.e. the (untwisted) Kac Moody
extension of $SL(2,{\bf R})$, with a central term acting as a
scaling operator on the conformal factor. These results suggested
further links between Einstein's theory and generalized Kac Moody
algebras, as well as the emergence of yet bigger symmetries
in the dimensional reduction. The results presented here
indicate that, if such extensions of the Geroch group exist, they
are likely to involve the topological degrees of freedom.
Stated in more physical terms, we are looking
for ``solution generating symmetries" that not only relate solutions
of the same topological type and with the same conformal structure
of the world sheet (e.g. asymptotically flat solutions of Einstein's
equations), but symmetries that permit changes of the topology and the
conformal structure as well.

The models considered here are most conveniently derived from
matter coupled supergravity theories in three dimensions,
i.e. locally supersmmetric non-linear sigma models as recently
formulated in \cite{dWNT}. This procedure has the advantage
that in three dimensions, all finite dimensional symmetries
are manifest because the matter degrees of freedom are uniformly
represented by scalars and spinors rather than tensor
fields (as would be the case in dimensions $d>3$).
The models obtained in the reduction to two dimensions
resemble conformal field theories in several respects, but there are
also differences. For instance, the equations of motion
of the left and right moving degrees of freedom can no longer
be disentangled, because there exist genuine solitonic solutions
mixing left and right movers such as the gravitational ``colliding
plane wave" solutions of \cite{KhP} also considered in \cite{Schlad}.
Furthermore, locally supersymmetric theories exist up to $N=16$
(where $N$ is the number of local supersymmetries),
whereas in conformal supergravity, only $N\leq 4$ is possible.
The difference is perhaps more clearly understood by a glance at
the (super)gravitational fields, which do not carry propagating
degrees of freedom. The bosonic ones originate from the dreibein in
three dimensions, which by a partial gauge fixing of the Lorentz
group $SO(1,2)$ can be cast into the form\footnote{We will use
letters $m,n,...$ and $a,b,...=0,1,2$, respectively, for curved
and flat indices in three dimensions, while the corresponding
indices in two dimensions will be denoted by $\mu ,\nu ,...$ and
$\a , \b ,...=0,1$, respectively. The metric has signature
$(+--)$. }
\beq
\euo ma = \pmatrix{ \euo \mu \a & \r A_\mu  \cr
                     0   & \r  \cr}  \Longrightarrow
\euo am = \pmatrix{ \euo \a \mu & - \euo \a \nu A_\nu  \cr
                     0   & \r^{-1}  \cr}
\labelx{dreibein}
\eeq
For the $3d$ gravitino, we have an analogous
decomposition in terms of flat indices
\beq
\p_a = (\p_\a , \p_2)
\labelx{3d gravitino}
\eeq
Dimensional reduction to two dimensions therefore gives rise to a
``dilaton" $\r$ and a Kaluza-Klein vector field $A_\mu$ in addition to
the zweibein $\euo \mu \a$, which is the only gravitational degree of
freedom in conformal field theory\footnote{We note that
a similar dilaton field has been considered recently in the context
of $2d$ conformal field theory \cite{Chams} and black hole
models \cite{CGHS}. However, there it is put in ``by hand" and
governed by a different Lagrangian.}. Similarly, the decomposition
\rf{3d gravitino} gives rise to an extra degree of freedom,
namely the ``dilatino" $\p_2$, which may be viewed as the superpartner
of $\r$. None of these fields possesses propagating degrees
of freedom. For the bosonic fields, this can be seen by
substituting \rf{dreibein} into the $3d$ Einstein action
and discarding the dependence on the third (spacelike)
coordinate $x^2$, which yields
\beq
- \ft14 e^{(3)} R^{(3)} = - \ft14 \r e R^{(2)}
       - \ft1{16} e \r^3 A_{\mu \nu} A^{\mu \nu}
\labelx{3d Einstein action}
\eeq
with $A_{\mu \nu} := \pa_\mu A_\nu - \pa_\nu A_\mu $.
Evidently, the conformal factor does not decouple
even in the classical theory as the Euler density $eR^{(2)}$
is multiplied by the dilaton field $\r$. Instead, there
is now an equation of motion relating the world sheet
curvature to matter sources.
The field $\r$ can be identified with a function of the coordinates
(for axisymmetric stationary solutions of Einstein's equations, it is
usually taken to be a cylindrical coordinate, see \cite{Hoens, BM}).
Nevertheless, it modifies the dynamics of the matter fields through
its appearance in their equations of motion. It also plays an
essential role in establishing one-loop finiteness of the
dimensionally reduced models \cite{dWGNR}. The vector field
$A_\mu$ is auxiliary, but offers the possibility of introducing
a cosmological constant through a non-vanishing expectation
value $A_{\mu \nu} \propto \e_{\mu \nu}$.
In previous work, this cosmological constant has always been
assumed to vanish, and we will also set it equal to zero here.
Elimination of the field strength $A_{\mu \nu}$ will then produce
only quartic spinor terms, which we do not consider in
this paper, so effectively $A_{\mu \nu} =0$. However, it must be
emphasized that inclusion of the associated field equation into
the linear system, which has so far not been accomplished, may
provide the crucial missing link in understanding the hidden
symmetries that may exist beyond the Geroch group.

As already mentioned, previous studies are based on the special
superconformal gauge
\beq
{e_\mu}^\a = \l \, \delta_\mu^\a    \;\;\; ,\;\;\;
\p_\a = \g_\a \t
\labelx{superconformal gauge}
\eeq
which simplifies the equations of motion considerably. This gauge
choice is always possible {\it locally}, but it misses important global
aspects because the information about the conformal structure of the
world sheet is hidden in the transition functions between local
charts in this gauge. Consequently, a change of conformal structure
must be accompanied by a corresponding change of atlas if
\rf{superconformal gauge} is to be maintained.
If one wants to vary the conformal structure without having to
change the atlas, one must make the dependence on the topological
degrees of freedom explicit. In order to do so,
one parametrizes the conformal structure over a fixed
atlas in terms of Beltrami and super-Beltrami differentials.
In the context of conformal field theory and string theory,
such a formulation was proposed in \cite{BB} and further
investigated in \cite{Dick, Stora}; it was also used in studies
of higher loop amplitudes in superstring theory \cite{AMS}.
In section 2, we will further develop this formalism, mainly
relying on and extending the results of \cite{Dick}, and use it in
the construction of the linear system in section 4.

Although our results could be formulated in the Euclidean metric
relevant to the study of stationary axisymmetric solutions,
we will be working with a Lorentzian worldsheet in this paper.
A technical reason for this is the occurrence of Majorana Weyl
spinors in two dimensions, which are here described as real
one component (anticommuting) spinors. As is well known,
Majorana Weyl spinors in two dimensions exist only for Lorentzian
signature, but not for Euclidean signature. This does not
necessarily preclude a Euclidean description, which would require
complex spinors. However, by complexifying the spinors, one doubles the
number of fermionic degrees of freedom. In a theory with an even
number of fermions, this problem can be circumvented by
rewriting $d$ real spinors in terms of $d\over 2$ complex spinors,
but some of the previously manifest symmetries would be lost in
general. Quite apart from these technical points, however,
the study of Lorentzian world sheets is of
interest in its own right. These differ from the more familiar
Euclidean world sheets (Riemann surfaces) in various respects,
one of which is the unavoidability of singularities for higher genus
surfaces: a globally Lorentzian surface which is everywhere
smooth must have Euler characteristic $\chi = 2-2g-n=0$
\cite{Steenrod} (where $g$ is the genus
and $n$ the number of punctures). This leaves only the cylinder
($g=0, n=2$) and the torus ($g=1, n=0$) as everywhere smooth Lorentzian
world sheets, so all other worldsheets must have singularities.
These observations are also of some physical interest, for instance
in two dimensional quantum cosmology (see e.g.\cite{Susskind}),
where they imply the existence of catastrophic ``naked" singularities
for two dimensional observers\footnote{A nice realization of the
higher genus surfaces is provided by the Mandelstam
diagrams of closed string theory \cite {GW}. These are smooth
(in fact, flat) surfaces except at the points where two strings
join and the curvature is proportional to a delta function.}.
Another peculiar feature of genus one (and possibly higher genus)
Lorentzian world sheets has been stressed recently in \cite{Moore}:
the modular group acts ergodically on Teichm\"uller
space (but can also have periodic orbits for non-generic
points!) and thus the quotient of this space
by the modular group gives rise to a very strange moduli space.
Unfortunately, owing to the lack of literature
dealing with the geometry of ``Lorentzian Riemann surfaces"
from either a mathematical or a physical point of view,
many elementary questions remain open for the time being.
We will proceed nonetheless, assuming that the known results
about ordinary Riemann surfaces can be taken over mutatis mutandis.

\Section{Conformal Calculus for Lorentzian Worldsheets}

In this section, we consider world sheets which are globally
Lorentzian two-dimensional manifolds, possibly with singular points
as we already discussed. The local charts are parametrized by conformal
(i.e. light cone) coordinates $(x^\dop , x^\dom )$ \footnote{Light-cone
components are defined by $V^\pm := {1\over {\sqrt{2}}}
(V^0 \pm V^1 )$.}.
To distinguish flat (Lorentz) from curved (world) indices, we will
put dots on the latter. Inequivalent world sheets of the same
topological type are classified by their conformal structure
(i.e. the complex structure for Euclidean Riemann surfaces).
As already mentioned in the introduction, there are
two ways to describe them. One can either choose conformal
coordinates, i.e. a diagonal (``conformal") gauge for the zweibein,
or otherwise parametrize the conformal structure by Beltrami
differentials. The first option corresponds to the standard
description of conformal field theories \cite{Friedan, FMS}.
However, we here prefer to make use of the second possibility, which
has the advantage that one can keep the atlas and the transition
functions fixed while varying the conformal structure
\cite{BB, Dick}. Accordingly, we parametrize the zweibein as
\beq
\euo \mu \a = \pmatrix{ \epp & \mpm \emm   \cr
                        \mmp \epp & \emm}
\labelx{zweibein}
\eeq
where $\mpm$ and $\mmp$ are the Beltrami differentials, subject to
the condition $\mpm \mmp < 1$ (for Euclidean signature,
they are each other's complex conjugates, but here they are two
independent real fields).
The metric is given by $g_{\mu \n} = {e_\mu}^\a {e_\n}^\b
\eta_{\a \b}$, where $\eta_{+-}=1 , \eta_{++} = \eta_{--} = 0$.
With this parametrization, the line element assumes the form
\beq
ds^2 = 2 \epp \emm \big( \dxp + \mmp \dxm \big)
\big( \dxm + \mpm \dxp \big)
\labelx{lineelement}
\eeq
{}From its invariance, it is straightforward to determine the
transformation properties of the Beltrami differentials and the
prefactor $\epp \emm$ under general coordinate transformations
$x^\dop \rightarrow {x'}^\dop (x^\dop , x^\dom ), x^\dom
\rightarrow {x'}^\dom (x^\dop , x^\dom )$. Putting primes
on all terms on the right hand side and requiring the
primed and unprimed expressions to be equal, we obtain
\beq
\epp = {e'_\dop}^+ \big( \pa_\dop {x'}^\dop + {\mu'_\dom}^\dop
       \pa_\dop {x'}^\dom \big) \;\;\; , \;\;\;
\emm = {e'_\dom}^- \big( \pa_\dom {x'}^\dom + {\mu'_\dop}^\dom
       \pa_\dom {x'}^\dop \big)
\eeq
The Beltrami differentials transform as
\beq
\mpm =
{{\pa_\dop {x'}^\dom + {\mu'_\dop}^\dom \pa_\dop {x'}^\dop }\over
 {\pa_\dom {x'}^\dom + {\mu'_\dop}^\dom \pa_\dom {x'}^\dop }}
      \;\;\; , \;\;\;
\mmp =
{{\pa_\dom {x'}^\dop + {\mu'_\dom}^\dop \pa_\dom {x'}^\dom }\over
 {\pa_\dop {x'}^\dop + {\mu'_\dom}^\dop \pa_\dop {x'}^\dom }}
\eeq
The inverse formulas read
\beq
{\mu'_\dop}^\dom =
{{ \mpm \pa_\dom {x'}^\dom - \pa_\dop {x'}^\dom} \over
{\pa_\dop {x'}^\dop - \mpm \pa_\dom {x'}^\dop }}  \;\;\; , \;\;\;
{\mu'_\dom}^\dop =
{{ \mmp \pa_\dop {x'}^\dop - \pa_\dom {x'}^\dop} \over
{\pa_\dom {x'}^\dom - \mmp \pa_\dop {x'}^\dom }}
\labelx{Beltrami trafo}
\eeq
To make the factorization of two dimensional diffeomorphisms
manifest, we now switch to an anholonomic basis for the
derivatives and the differentials, following \cite{Dick}
which is based on but differs from earlier work in \cite{BB}.
For the derivative operators, we define
\beq
\Dp := \pa_\dop - \mpm \pa_\dom    \;\;\; ,\;\;\;
\Dm := \pa_\dom - \mmp \pa_\dop
\labelx{Dvector}
\eeq
In terms of these, left and right moving scalar fields
satisfy $\Dm f =0$ and $\Dp \bar f =0$, respectively; they are the
real analogues of holomorphic and anti-holomorphic functions.
The dual basis differential forms are then
\beq
\Dxp := {{\dxp + \mmp \dxm}\over {\MM }}  \;\;\; ,\;\;\;
\Dxm := {{\dxm + \mpm \dxp}\over {\MM }}  \;\;\; ,\;\;\;
\labelx{Dx}
\eeq
It is important here that the factor $\MM$ is assigned to the
differential forms rather than the derivatives;
for any other assignment, the factorization of $2d$
diffeomorphisms does not work \cite{Dick}.
With these definitions, \rf{Dvector} and \rf{Dx}
transform as follows under general coordinate transformations
\beql
\Dp &=& \Dp {x'}^\dop \, {\cal D}'_\dop  \;\;\; , \;\;\;
\Dm  =  \Dm {x'}^\dom \, {\cal D}'_\dom  \zeile
{\cal D} {x'}^\dop &=& \Dp {x'}^\dop \, \Dxp   \;\;\; , \;\;\;
{\cal D} {x'}^\dom  =  \Dm {x'}^\dom \, \Dxm
\labelx{2d Diff}
\eeql
while \rf{Beltrami trafo} takes the simple form
\beq
{\mu'_\dop}^\dom =
- {{\Dp {x'}^\dom}\over {\Dp {x'}^\dop}} \;\;\; ,\;\;\;
{\mu'_\dom}^\dop =
- {{\Dm {x'}^\dop}\over {\Dm {x'}^\dom}}
\labelx{mu prime}
\eeq
Use of \rf{2d Diff} and \rf{mu prime} and a little further
algebra show that
\beq
\Dp {x'}^\dop  = \left( {\cal D}'_\dop x^\dop \right)^{-1} \;\;\; ,
\;\;\; \Dm {x'}^\dom  = \left( {\cal D}'_\dom x^\dom \right)^{-1}
\labelx{Factor Trafo}
\eeq
Since these relations are valid for arbitrary diffeomorphisms
$x^\mu \rightarrow {x'}^\mu (x)$, the factorization of
$2d$ diffeomorphisms is now completely explicit.
The volume element is
\beq
\Dxp \wedge \Dxm = {{\dxp \wedge \dxm}\over {\MM}}
\eeq
so that e.g. for a scalar field $\ve$, we have
\beq
\int d^2 x \sqrt{g} g^{\mu \nu} \pa_\mu \ve \pa_\nu \ve =
2\int \Dxp \wedge \Dxm \, \Dp \ve \Dm \ve
\eeq

Rather than the zweibein components $\epp$ and $\emm$, which do
not transform properly, we must use the chiral ``einbeine"
\beq
\tepp := \epp (\MM ) \;\;\; , \;\;\; \temm := \emm (\MM )
\labelx{epmtilde}
\eeq
and their inverses
\beq
\itepp := \left( \tepp \right)^{-1} \;\;\; , \;\;\;
\itemm := \left( \temm \right)^{-1}
\labelx{epminvtilde}
\eeq
by means of which flat chiral indices $\pm$ can be converted
into curved chiral indices $\dot \pm$ and vice versa. Note that
$\itepp$ and $\itemm$ are just the diagonal entries of the
inverse zweibein and that there are no
off-diagonal components $\tilde e_\dop^{\;\; -}$ and
$\tilde e_\dom^{\;\; +}$ in this formalism!
Under diffeomorphisms, \rf{epmtilde} and \rf{epminvtilde} transform
in the same manner as the derivatives ${\cal D}_{\dot \pm}$
and the differentials ${\cal D} x^{\dot \pm}$, respectively;
under the local Lorentz group $SO(1,1)$, $\tepp$ and $\temm$ scale
oppositely. We also define the Lorentz scalar
\beq
\te := \tepp \temm = \epp \emm (\MM )^2
\labelx{etilde}
\eeq
Apart from the $\mu$-dependent modifications required for the proper
behavior under reparametrizations, $\te$ is just the (square of
the) conformal factor.

As in ordinary complex calculus \cite{Friedan}, we can refer all
tensors to the basis \rf{Dx}; in analogy with the tensor
calculus on ordinary Riemann surfaces, we will then call them
``differentials", or primary fields. A differential $T$ is
consequently defined by requiring
\beq
T \equiv \cT (x^\dop, x^\dom ) (\Dxp )^m (\Dxm )^n
\labelx{differentials}
\eeq
to be invariant under coordinate transformations (here and in the
sequel, we use script letters for differentials defined with
respect to the anholonomic basis \rf{Dx}). Alternatively,
we can define $\cT$ by converting the corresponding tensor with
flat indices into one with curved chiral indices by means of
\rf{epmtilde} and \rf{epminvtilde}. The pair $(m,n)$ is the
degree or ``conformal weight" of $T$; as we will see below,
$m,n$ can be integer or half-integer. For instance,
$\tepp$ and $\temm$ are differentials of degree $(1,0)$ and
$(0,1)$, respectively, whereas the Beltrami ``differentials"
are not proper tensors as is obvious from (2.9).

In order to define covariant derivatives, we must introduce the
appropriate Christoffel symbols. Let us first determine the
coefficients of anholonomy defined by
$\O_{\a \b \g} := \euo \a \mu \euo \b \n (\pa_\mu e_{\n \g} -
\pa_\n e_{\mu \g} )$. With the zweibein parametrized as in
\rf{zweibein}, we have\footnote{With $\iepp := (\epp )^{-1}$ and
$\iemm := (\emm )^{-1}$.}
\beql
\O_{-+-} &=& {{ \iepp \iemm} \over \MM} \, \left( \pa_\dom \epp
 -\pa_\dop (\mmp \epp ) \right)  \zeile &=&   {1\over \MM}
 \, \iemm \left( \iepp \Dm \epp - \pa_\dop \mmp \right) \zeile
\O_{+-+} &=& {{\iepp \iemm } \over \MM} \,  \left( \pa_\dop \emm
 -\pa_\dom (\mpm \emm ) \right)  \zeile &=&
{1\over \MM} \, \iepp \left( \iemm \Dp \emm - \pa_\dom \mpm \right)
\eeql
{}From (A.6) in the appendix, we then get
$\o_{-+-} = \O_{-+-}$ and $\o_{+-+} = \O_{+-+}$. After a little
rearrangement, we can express the spin connection in terms
of the einbein fields $\tepp$ and $\temm$ as
\beq
\o_{+-+} = \itepp \left( \itemm \Dp \temm - \Gpmm \right) \;\;\; , \;
\;\; \o_{-+-} = \itemm \left( \itepp \Dm \tepp - \Gmpp \right)
\labelx{2d spin connection}
\eeq
where the new Christoffel symbols are defined by \cite{Dick}
\beq
\Gpmm :=
{{\Dm \mpm - \mpm \Dp \mmp}\over \MM} \;\;\; , \;\;\;
\Gmpp :=
{{\Dp \mmp - \mmp \Dm \mpm}\over \MM}
\labelx{Christoffel}
\eeq
Observe that they depend on the zweibein \rf{zweibein} only
through the Beltrami differentials. Readers should be warned that
these Christoffel symbols do {\it not} coincide with the
usual ones that one would compute from $\G_{\mu \nu}^\r$.
Similar remarks apply to the spin connection components with
curved chiral indices $\dot \pm$ to be defined below, which
are not the same as the ones computed from $\o_{\mu \a \b}$.
(On the other hand, all quantities with flat indices are the
same as in the usual formalism!) To make this distinction completely
explicit, we put tildes on all chiral tensors that differ from
the usual ones. The chiral einbeine $\tepp$ and $\temm$ obey a
factorized version of the usual vielbein postulate, viz.
\beql
\Dp \temm + \opmp \temm = \Gpmm \temm  \;\;\; &,& \;\;\;
\Dp \tepp + \oppm \tepp = \Gppp \tepp   \zeile
\Dm \tepp + \ompm \tepp = \Gmpp \tepp  \;\;\; &,& \;\;\;
\Dm \temm + \ommp \temm = \Gmmm \temm   \zeile
\labelx{vielbein postulate}
\eeql
with
\beq
\tilde \o_{\dop -+} := \tepp \o_{+-+}   \;\;\; , \;\;\;
\tilde \o_{\dom +-} := \temm \o_{-+-}
\eeq
and
\beq
\Gppp := \te^{-1} \Dp \te \; - \; \Gpmm\;\;\; , \;\;\;
\Gmmm := \te^{-1} \Dm \te \; - \; \Gmpp
\labelx{Christoffel2}
\eeq
Note the absence of components $\Gppm$ and $\Gmmp$
in this formalism; the Christoffel
symbol thus has only four distinct components instead
of the usual eight. For $\mpm = \mmp = 0$, we recover
the usual formulas of conformal (complex) tensor calculus
\cite{Friedan}. For completeness, let us also list the commutator
of $\Dp$ and $\Dm$,
\beq
\big[ \Dp \, , \, \Dm \big] = \Gpmm \Dm - \Gmpp \Dp
\labelx{Kommutator curly D}
\eeq
This means that for $\mpm , \mmp \neq 0$ there is ``torsion",

We now define covariant derivatives (denoted by straight Roman
letters $\rD_{\dot \pm}$) on arbitrary $(m,n)$-differentials $\cT$
\beql
\rD_\dop \cT &:=& \Dp \cT - m \Gppp \cT - n \Gpmm \cT  \zeile
\rD_\dom \cT &:=& \Dm \cT - m \Gmpp \cT - n \Gmmm \cT
\labelx{covariant derivative1}
\eeql
Likewise, we can define covariant derivatives on mixed tensors by
use of the spin connection and the Christoffel symbols.
\rf{vielbein postulate} shows that the conversion of flat chiral
indices into curved ones by means of \rf{epmtilde} and
\rf{epminvtilde} is a covariant operation. From \rf{vielbein postulate}
it also follows that the (1,1) density $\te$
(cf. \rf{etilde}) is covariantly constant, i.e.
$\rD_\dop \te = \rD_\dom \te = 0$. This is the analog of the
covariant constancy of the metric tensor in the usual
tensor formalism. Evaluating the commutator of two
covariant derivatives on a $(m,n)$ differential, we obtain
\beq
\big[ \rD_\dop \, ,\, \rD_\dom \big] = (-m+n) \,  \cR
\eeq
where the curvature $\cR$ is defined by
\beql
\cR &:=&  \Dp \Gmpp - \Dm \Gppp + \Gmpp \Gppp - \Gpmm \Gmpp   \zeile
   &=& \Dm \Gpmm - \Dp \Gmmm + \Gmpp \Gmmm - \Gmpp \Gpmm
\labelx{2d curvature}
\eeql
It is a (1,1) differential, and related to the usual scalar
curvature by $R^{(2)} = 2\itepp \itemm \cR $, where
\beql
R^{(2)} &=& -2 e^{-1} \pa_\mu \big( e e^{\a \mu} {\O_{\a \b}}^\b \big)
           \zeile &=& -2
e^{-1} \pa_\mu \Big( {e_+}^\mu \pa_\nu (e {e_-}^\nu ) + {e_-}^\mu
  \pa_\nu (e {e_+}^\nu )\Big)
\eeql

As for Euclidean worldsheets, one can define half order
differentials required for the description of fermions \cite{GS}
by multiplying the chiral spinor components with
appropriate half-integer powers of $\tepp$ and $\temm$.
The half order differentials are inert with respect to
local Lorentz transformations, and transform with half-integer
powers of $\Dp {x'}^\dop$ and $\Dm {x'}^\dom$ under general
coordinate transformations. For Majorana Weyl spinors,
the chiral components are real and have only one (anticommuting)
component\footnote{For Euclidean signature, the chiral spinor
components are complex because the local Lorentz group $SO(2)$ acts
on them by a $U(1)$ phase transformation.}.
The Lorentz covariant derivative on a spinor $\chi$ is given by
\beq
\rD_\a \chi = \left( \pa_\a - \ft12 \o_{\a -+}
    \g^3  \right) \chi
\labelx{Dslash1}
\eeq
(see the appendix for our $\g$-matrix conventions; as before,
we use straight letters $\rD_\mu$ to denote gravitationally
and/or Lorentz covariant derivatives). By means of the formulas
\rf{epmtilde}, \rf{epminvtilde} and \rf{2d spin connection}
above, we can evaluate the derivative on the chiral components
of $\chi$. Setting $\a = \pm$, we find
\beql
\rD_+ \chi_+ &=& (\itepp )^{3/2} \left( \Dp - \ft12 \Gppp \right) \chi_\dop
              =  (\itepp )^{3/2} \rDp \chi_\dop     \zeile
\rD_- \chi_+ &=& \itemm (\itepp )^{1/2} \left( \Dm - \ft12 \Gmpp \right)
   \chi_\dop = \itemm (\itepp )^{1/2} \rDm \chi_\dop
\labelx{Dpm chi}
\eeql
where
\beq
\chi_\dop := (\tepp )^{1/2} \chi_+  \;\;\; , \;\;\;
\chi_\dom := (\temm )^{1/2} \chi_-
\eeq
\rf{Dpm chi} are the properly covariantized derivatives for
a $(\ft12 , 0)$ differential. The evaluation of the derivatives
$\rD_\pm$ on the negative chirality component $\chi_-$ works
in exactly the same way.
The redefinition of Lorentz spinors by square roots of the
chiral zweibein components is the same as in Euclidean conformal
field theory, but the dependence on the Beltrami differentials
has so far not been exhibited as previous work has relied on the
conformal gauge.

The decomposition of the gravitinos into differentials is slightly
more involved. Making use of the split \rf{3d gravitino},
the dilatino component $\p_2$ can be converted into a pair of
$(\halb ,0)$ and $(0,\halb )$ differentials as $\chi$ above.
On the other hand, $\p_\a$ must first be
decomposed into irreducible components according to
$\p_\a = \tilde \p_\a + \g_\a \t$ , where $\g^\a \tilde \p_\a =0$.
Then $(\p_+)_+ = (\tilde \p_+)_+$, $(\p_+)_- = \g_+ \t_+$ and
$(\p_-)_- = (\tilde \p_-)_-$, $(\p_-)_+ = \g_- \t_-$.
The super Beltrami differentials are defined as
\beq
{\p_\dop}^\dom := \tepp (\itemm)^{{1\over 2}} (\p_+)_+  \;\;\; , \;\;\;
{\p_\dom}^\dop := \temm (\itepp)^{{1\over 2}} (\p_-)_-
\labelx{superBeltrami}
\eeq
They thus have conformal weight $(1,-\ft12 )$ and $(-\ft12 ,1)$,
respectively. That \rf{superBeltrami} is the the correct definition
can be seen from the dimensional reduction of the Rarita
Schwinger equation (see also the following section).
For instance, a little calculation which is
completely analogous to \rf{Dpm chi} shows that
\beq
\rD_+  (\p_-)_- = \itemm (\tepp )^{1\over 2} \rDp {\p_\dom}^\dop
\eeq
where $\rD_+$ on the left hand side is the Lorentz covariant derivative,
while $\rDp$ on the right hand side is the covariant derivative
\rf{covariant derivative1} with $(m,n)=(1,-\ft12 )$.

Finally, the supersymmetry transformation parameters turn
turn out to be half order differentials of weight $(-\ft12 ,0$ and
$(0,-\halb )$, respectively, and are defined by
\beq
\e^\dop := (\itepp )^{{1\over 2}} \e_+  \;\;\; ,\;\;\;
\e^\dom := (\itemm )^{{1\over 2}} \e_-
\eeq

\Section{Equations of Motion and Dimensional Reduction}

We will now list the equations of motion in three dimensions and
reduce them to two dimensions. For notational
simplicity, we will write down the formulas for $N=16$ supergravity
\cite{MS} only, the generalization to other $N$ being straightforward
(see \cite{dWNT} for a comprehensive discussion of these models).
Our conventions and notation are the same as in \cite{MS, NW},
and we therefore summarize them only briefly. The model is a locally
supersymmetric sigma model based on the non-compact coset space
$E_{8(+8)}/SO(16)$. The $E_8$ generators are decomposed into the
120 generators $X^{IJ}=-X^{JI}$  of the $SO(16)$ subgroup
and 128 remaining generators $Y^A$, which transform as the irreducible
spinor representation of $SO(16)$. Thus $I,J,...=
1,...,16$ are $SO(16)$ vector indices and $A,B,...=1,...,128$
are $SO(16)$ spinor indices. The matter fermions $\chi^\Adot$
transform under the conjugate spinor representation labeled by
dotted indices $\Adot , \Bdot , ... = 1,...,128$. The bosonic sector
of the $N=16$ theory is governed by a non-linear sigma model; thus,
the bosonic fields are described by a matrix $\V (x) \in E_8$, which
is subject to the transformations
\beq
\V (x) \longrightarrow g^{-1} \V (x) h(x)
\eeq
where $g$ is a rigid $E_8$ transformation, and $h(x)$ a local
$SO(16)$ transformation. From $\V$, one defines the ``composite fields"
$Q_m^{IJ}$ and $P_m^A$
\beq
\V^{-1} \pa_m \V = \halb Q_m^{IJ} X^{IJ} + P_m^A Y^A
\labelx{3d VdV}
\eeq
This definition immediately implies the integrability relations
\beql
\rD_m P_n^A - \rD_n P_m^A &=& 0 \zeile
\pa_m Q_n^{IJ} -\pa_n Q_m^{IJ} + 2 Q_m^{K[I} Q_n^{J]K} &+&
      \ft12 \GIJ AB P_m^A P_n^B  = 0
\labelx{3d integrability}
\eeql
where the $SO(16)$ covariant derivative $\rD_m$ is defined by means
of the connection $Q_m^{IJ}$ defined in \rf{3d VdV}.
Rather than write down the Lagrangian (see \cite{MS}), we will
give the equations of motion right away, disregarding higher order
fermionic terms. The Rarita Schwinger equation
for the 16 gravitinos $\p_m^I (x)$ is
\beq
\e^{mnp} \rD_n \p_p^I = \ft12 \g^n \g^m \chi^\Adot \GI A \Adot P_n^A
\labelx{3d RS}
\eeq
where the Lorentz and $SO(16)$ covariant derivative is defined by
\beq
\rD_m \p^I_n := \Big( \delta^{IJ} \big( \pa_m + \ft14 \o_{mab}
   \g^{ab} \big) + Q_m^{IJ} \Big) \p_n^J
\eeq
The 128 matter fermions $\chi^\Adot$ are subject to
\beq
-i\g^m \rD_m \chi^\Adot = \ft12 \g^n \g^m \p_n^I P_m^A \GI A \Adot
\labelx{3d Dirac}
\eeq
with
\beq
\rD_m \chi^\Adot := \left( \delta^{\Adot \Bdot} \left( \pa_m +
      \ft14 \o_{mab} \g^{ab}\right) + \ft14 Q_m^{IJ}
      \GIJ \Adot \Bdot \right) \chi^\Bdot
\eeq
The scalar field equation reads
\beql
D^m \Big( P_m^A &-& \GI A \Adot \bar \chi^\Adot \g^n \g_m \p_n^I
                      \Big) =  \zeile
 &=&   \ft12 \e^{mnp} \bar \p_m^I \p_n^J \GIJ AB P_p^B +
     \ft18 i \bar \chi \g^m \G^{IJ} \chi \, \GIJ AB P_m^B
\labelx{3d scalar}
\eeql
Variation of the dreibein leads to Einstein's equation
\beql
R_{ma} &-& \ft12 e_{ma} R = P_m^A P_a^A -
           \ft12 e_{ma} g^{pq} P_p^A P_q^A   \zeile
    &&  - \,   i\bar \chi^\Adot \g_a D_m \chi^\Adot  +
  e_{ma}  i\bar \chi^\Adot \g^p D_p \chi^\Adot  \zeile
&& + \, \GI A \Adot \left( e_{ma} \bar \chi^\Adot \g^p \g^q \p_p^I P_q^A
 - \bar \chi^\Adot \g_a \g^p \p_m^I P_p^A
 - \bar \chi^\Adot \g^p \g_a \p_p^I P_m^A \right)  \zeile
\labelx{3d Einstein}
\eeql
This expression is not symmetric under interchange of $m$ and $a$
in first order formalism. However, the right hand side can be
rendered symmetrical by substituting
the second order spin connection
\beq
\widehat \o_{abc} = \o_{abc} (e)
        + \ft14 \e_{abc} \bar \chi^\Adot \chi^\Adot
   + \ft12 i \bar \p_b^I \g_a \p_c^I
   + i \bar \p^I_{\lbrack b} \g_{c \rbrack} \p_a^I
\eeq
into the Einstein tensor and shifting the resulting fermionic
terms from the left to the right hand side. After a little
calculation, making use of the fermionic field equations
\rf{3d RS} and \rf{3d Dirac}, one arrives at the symmetric result
\beql
R_{ab} (e) &-& \ft12 \eta_{ab} R (e) = P_a^A P_b^A -
     i\bar \chi^\Adot \g_{(a} D_{b)} \chi^\Adot  \zeile
&+&  \eta_{ab} \left( i \bar \chi^\Adot \g^m D_m \chi^\Adot
    - \ft12  g^{mn} P_m^A P_n^A  \right)  \zeile
&+& \GI A \Adot \left( \eta_{ab} \bar \chi^\Adot \g^p \g^q \p_p^I P_q^A
 - \bar \chi^\Adot \g_{(a} \g^p \p_{b)}^I P_p^A
 - \bar \chi^\Adot \g^p \g_{(a} \p_p^I P_{b)}^A \right)\zeile
&+& D^m \left( i\bar \p^I_{(a} \g_{b)} \p_m^I \right) -
D_{(a} \left( i\bar \p^I_{b)} \g^c \p_c^I \right)
+ \eta_{ab} D^m \Big( i\bar \p_m^I \g^c \p_c^I \Big)
\labelx{3d symmetric Einstein}
\eeql
where $(...)$ denotes symmetrization with strength one.
Contracting with $\eta^{ab}$, we obtain
\beq
R^{(3)} = g^{mn} P_m^A P_n^A -
           2i D^m \left( \bar \p_m^I \g^n \p_n^I \right)
\labelx{3d R}
\eeq
where \rf{3d Dirac} has been used again.

Modulo higher order fermionic terms, these equations are covariant
with respect to the local supersymmetry variations
\beql
 \d \euo ma &=& i\bar \e^I \g^a \p_m^I \zeile
 \d \p_m^I &=& \rD_m \e^I    \zeile
 \d \chi^\Adot &=& \ft12 i \g^m \e^I \GI A \Adot P_m^A  \zeile
 \Vinv \d \V  &=& \GI A \Adot \bar \e^I \chi^\Adot \, Y^A
\labelx{3d supervariations}
\eeql

To reduce these equations to two dimensions we drop the
dependence on the third spacelike coordinate $x^2$,
using the decompositions \rf{dreibein} and \rf{3d gravitino}.
We then rewrite all equations in terms of the chiral basis
introduced in the foregoing section, making use of the conformal
calculus developed there. For this purpose, we need the
coefficients of anholonomy $\O_{abc}$ in the basis \rf{dreibein}
\beql
\O_{\a \b \g} &=& 2 \euo {[\a} \mu \euo {\b ]} \nu
          \pa_\mu e_{\nu \g}     \zeile
\O_{\a \b 2} &=& - \r \euo \a \mu \euo \b \nu A_{\mu \nu }
\;\;\; , \;\;\; \O_{2 \b \g} = 0     \zeile
\O_{\a 2 2} &=& - \euo {\a} \mu \r^{-1} \pa_\mu \r
\eeql
(Remember that, with our metric ${\O_{\a \b}}^2 = - \O_{\a \b 2}$).
The first of these has already been evaluated in terms of the
anholonomic basis \rf{Dx} in \rf{2d spin connection}.
For the remaining components, we get
\beq
\o_{+-2} = - \o_{-+2} = - \o_{2+-} = \ft12 \O_{+-2}
\;\;\; , \;\;\; \o_{2 \pm 2} = \O_{2 \pm 2}
\labelx{omega+-2}
\eeq
with
\beql
\O_{+- 2} &=&  -\r \itepp \itemm \cA_{\dop \dom} \zeile
\O_{2+2} &=& \itepp \r^{-1} \Dp \r \;\;\; , \;\;\;
\O_{2-2} = \itemm \r^{-1} \Dm \r
\labelx{Omega+-2}
\eeql
where $\cA_{\dop \dom}$ is the Maxwell field strength
\beq
\cA_{\dop \dom} := \rD_\dop \cA_\dom - \rD_\dom \cA_\dop
\labelx{Maxwell}
\eeq
of the Kaluza Klein vector field $\cA_{\dot \pm}$ in the
``curly basis"
\beq
A_\mu dx^\mu = \cA_\dop \Dxp + \cA_\dom \Dxm
\eeq
(because of the non-vanishing torsion in \rf{Kommutator curly D}
the Christoffel symbols do not drop out in \rf{Maxwell}).
We can now compute the components of the $3d$ Riemann tensor
in this reduction. Using
\beq
R_{abcd} = {\O_{ab}}^e \o_{ecd} + \pa_a \o_{bcd} - \pa_b \o_{acd}
+ {\o_{ac}}^e \o_{ebd} - {\o_{bc}}^e \o_{ead}
\labelx{3d Riemann}
\eeq
we get
\beql
R_{+-+-} &=&  \itepp \itemm \Big( - \cR + \ft34 \r^2
     \te^{-1} \cA_{\dop \dom} \cA_{\dop \dom} \Big)    \zeile
R_{2-+-} &=& - \ft12 \itemm \itepp \itemm \r^{-2} \Dm \Big( \r^3
   \cA_{\dop \dom} \Big)  \zeile
R_{2+-+} &=& - \ft12 \itepp \itemm \itepp \r^{-2} \Dp \Big( \r^3
   \cA_{\dom \dop} \Big)  \zeile
R_{2+2-} &=& \r^{-1} \itepp \itemm \Big( \rDp \Dm \r - \ft14 \r^3
   \te^{-1} \cA_{\dop \dom} \cA_{\dop \dom} \Big)    \zeile
R_{2+2+} &=& \itepp \itepp \r^{-1} \rDp \Dp \r  \;\;\; , \;\;\;
          R_{2-2-} = \itemm \itemm \r^{-1} \rDm \Dm \r
\labelx{2d Riemann}
\eeql
where the $2d$ curvature $\cR$ has been defined in \rf{2d curvature}.
(The components that have not been listed simply follow from the
well known symmetry properties of the Riemann tensor.) As already
indicated in the introduction, we will set $\cA_{\dop \dom} =0$
in the remainder of this paper, because we have not yet found a way
to include the associated equation of motion into the linear
system to be constructed in the next section.

In writing down the dimensionally reduced field equations, we will
also use the ``curly basis" for the fields $Q_\mu^{IJ}$ and
$P_\mu^A$. Hence,
\beq
Q_\mu^{IJ} dx^\mu = \cQ_\dop^{IJ} \Dxp +
         \cQ_\dom^{IJ} \Dxm    \;\;\; , \;\;\;
P_\mu^A dx^\mu = \cP_\dop^A \Dxp + \cP_\dom^A \Dxm
\labelx{curly QP}
\eeq
The integrability conditions \rf{3d integrability} now read
\beql
\rD_\dop \cP_\dom^A - \rD_\dom \cP_\dop^A &=& 0 \zeile
\rDp \cQ_\dom^{IJ} -\rDm \cQ_\dop^{IJ}+2 \cQ_\dop^{K[I} \cQ_\dom^{J]K}
   &+& \ft12 \GIJ AB \cP_\dop^A \cP_\dom^B  = 0
\labelx{2d integrability}
\eeql
where $\rD_{\dot \pm}$ now always denotes the fully covariant
derivative with respect to both \rf{covariant derivative1} and
local $SO(16)$.

{}From \rf{omega+-2} and \rf{Omega+-2} it is evident that the
$3d$ Lorentz covariant derivatives will give extra terms beyond
the ones exhibited in \rf{Dslash1}. Since we assume $\O_{+-2} =0$,
we must, however, only watch out for terms with $\o_{2 \pm 2}$.
Otherwise, the dimensional reduction of the fermionic field
equations is rather straightforward: we simply rewrite them in terms
of flat chiral indices and then convert them by means of the formulas
in the foregoing section.
In this way, we can show that \rf{3d Dirac} becomes
\beql
- i \r^{-1/2} \rDp \left( \r^{1/2} \chm \Adot \right) &=&
 - \ft12 i  \GI A \Adot \pzm I \cP_\dop^A +
 \wurzelhalb \GI A \Adot \smpm I \cP_\dom^A    \zeile
- i \r^{-1/2} \rDm \left( \r^{1/2} \chp \Adot \right) &=&
 + \ft12 i  \GI A \Adot \pzp I \cP_\dom^A +
 \wurzelhalb \GI A \Adot \smmp I \cP_\dop^A
\labelx{2d Dirac}
\eeql
{}From \rf{3d scalar}, we derive the scalar equation of motion
\beql
\r^{-1} \rDm \Big( \r \big( \cP_\dop^A &-& i\wurzelzwei \GI A \Adot
       \chp \Adot \pzp I + 2 \GI A \Adot \chm \Adot
       \smpm I \big) \Big) + \zeile  +
\r^{-1} \rDp \Big( \r \big( \cP_\dom^A &-& i\wurzelzwei \GI A \Adot
       \chm \Adot \pzm I - 2 \GI A \Adot \chp \Adot
       \smmp I \big) \Big)  \; =  \zeile
= \, \ft18 i \GIJ AB \Big(&-& \wurzelzwei \cP_\dop^B \, \GIJ \Adot \Bdot
    \chm \Adot \chm \Bdot + \wurzelzwei \cP_\dom^B
    \GIJ \Adot \Bdot \chp \Adot \chp \Bdot \Big) \zeile
+ \,  \GIJ AB  \Big( \big( \wurzelzwei \pzp I \thp J &-&
       \pzm I \smpm J \big) \cP_\dom^B  \,  +  \,
   \big( \wurzelzwei \pzm I \thm J -
      \pzp I \smmp J \big) \cP_\dop^B \Big) \zeile
\labelx{2d scalar}
\eeql
Apart from the presence of the topological fields, these equations
differ from the equations of motion of the corresponding rigidly
supersymmetric flat space sigma models because of their dependence
on the dilaton $\r$ and its superpartner $\p_2^I$.

{}From \rf{3d RS}, we deduce the following equations,
\beql
\wurzelzwei \rDp \thm I - \rDm \smpm I  &=&
\ft1{\sqrt{2}} i \GI A \Adot \chm \Adot \cP_\dop^A  \zeile
\wurzelzwei \rDm \thp I - \rDp \smmp I  &=&
\ft1{\sqrt{2}} i \GI A \Adot \chp \Adot \cP_\dom^A
\labelx{2d RS1}
\eeql
and
\beql
\rDm (\r \pzp I ) &=& \ft1{\sqrt{2}} i \Dp \r \, \smmp I  \zeile
\rDp (\r \pzm I ) &=& - \ft1{\sqrt{2}} i \Dm \r \, \smpm I
\labelx{2d RS2}
\eeql
as well as the ``super-Virasoro conditions"
\beql
\cS_\dop^I &:=&
\rDp (\r \pzp I ) \, - \, i \Dp \r \, \thp I \, + \,  \r \GI A \Adot
           \chp \Adot \cP_\dop^A  \; = \; 0  \zeile
\cS_\dom^I &:=&
\rDm (\r \pzm I ) \, + \,  i \Dm \r \, \thm I \, - \, \r \GI A \Adot
           \chm \Adot \cP_\dom^A   \; = \;  0
\labelx{2d RS3}
\eeql
corresponding to the variation of the traceless gravitino
modes $\tilde \p^I_{\pm}$.
Apart from contributions involving the topological degrees of
freedom, $\r \p_2^I$ is thus a free  field.

In the gravitational sector, the Einstein equation \rf{3d Einstein}
gives rise to several equations after dimensional reduction.
{}From the 22-component of \rf{3d symmetric Einstein}, we get
\beql
R_{22} &=& -i \bar \chi^\Adot \g^a \rD_a \chi^\Adot
   - \GI A \Adot \left( \bar \chi^\Adot \g^a \g^b \p_a^I P_b^A
           + \bar \chi^\Adot \g_2 \g^a \p_2^I P_a^A \right) \zeile
  && + \rD^a \big( i \bar \p_2^I \g_2 \p_a^I \big)
     - \rD_2 \big( i \bar \p_2^I \g^a \p_a^I \big)
\eeql
Invoking \rf{3d Dirac} and \rf{3d RS}, we can rewrite this as
\beq
R_{22}  = - \e^{abc} \bar \p_a^I \rD_b \p_c^I
   - 2 {\e_2}^{bc} \bar \p_2^I \rD_b \p_c^I
     + \rD^a \big( i \bar \p_2^I \g_2 \p_a^I \big)
     - \rD_2 \big( i \bar \p_2^I \g^a \p_a^I \big)
\eeq
Splitting the $3d$ Lorentz indices $a,b,...$ into $\a , \b ,...
= \pm$ and 2, and keeping track of the terms with $\o_{2 \pm 2}$,
we arrive at
\beq
\rD_\dop \rD_\dom \r = - \rDp \big( \r \smmp I \pzp I \big) -
     \rDm \big( \r \smpm I \pzm I \big)
\labelx{box rho}
\eeq
Thus, $\r$ would be a free field without the contributions
from the super-Beltrami differentials, consistent with the fact
that its superpartners $\r \p_2^I$ would also be free for
vanishing super Beltrami differentials.

For the curvature scalar, a similar calculation and use of
\rf{3d R} together with \rf{box rho} leads to
\beql
\cR &=&  \cP_\dop^A \cP_\dom^A
    \, + \, 2i \rDp \big( \smmp I \thp I \big)
    \, - \, 2i \rDm \big( \smpm I \thm I \big)   \zeile
&& - \sqrt{2} \r^{-1} \rDp \big( \r \pzm I \thm I \big)
   - \sqrt{2} \r^{-1} \rDm \big( \r \pzp I \thp I \big)
\labelx{2d curly Einstein}
\eeql
The variation of the off diagonal components of the zweibein
corresponding to $R_{++}$ and $R_{--}$ gives the
``Virasoro conditions"
\beql
\cT_{\dop \dop} &:=&
\rD_\dop \rD_\dop \r \, +\, \r \cP_\dop^A \cP_\dop^A \, - \,
 i \wurzelzwei \r \chp \Adot \rD_\dop \chp \Adot \, - \,
 2 \GI A \Adot \r \smpm I \chm \Adot \cP_\dop^A   \zeile
 && - \, i \sqrt{2} \GI A \Adot \r \chp \Adot \pzp I \cP_\dop^A
  \, - \, 2\sqrt{2} \GI A \Adot \r \chp \Adot \thp I \cP_\dop^A  \zeile
 && + \, \rDp \Big( \r \big( \pzm I \smpm I  \, + \,
     \sqrt{2} \thp I \pzp I \big) \Big)    \zeile
\cT_{\dom \dom} &:=&
\rD_\dom \rD_\dom \r \, +\,  \r \cP_\dom^A \cP_\dom^A \, + \,
 i \wurzelzwei \r \chm \Adot \rD_\dom \chm \Adot \, + \,
 2 \GI A \Adot \r \smmp I \chp \Adot \cP_\dom^A   \zeile
 && - \, i \sqrt{2} \GI A \Adot \r \chm \Adot \pzm I \cP_\dom^A
  \, + \, 2\sqrt{2} \GI A \Adot \r \chm \Adot \thm I \cP_\dom^A  \zeile
 && + \,  \rDp \Big( - \r \big( \pzp I \smmp I +
     \sqrt{2} \thm I \pzm I \big) \Big)
\labelx{Virasoro}
\eeql
When written out by means of \rf{Christoffel2}, one sees that the
terms $\rD_{\dot \pm} \rD_{\dot \pm} \r$ contains a contribution
proportional to $\te^{-1} {\cal D}_{\dot \pm} \te \,
{\cal D}_{\dot \pm}\r$, a term explicitly exhibited in previous
work based on the conformal gauge \rf{superconformal gauge},
see \cite{BM, Schlad}.
The terms $\rD_{\dot \pm} \rD_{\dot \pm} \r$ can also be expressed
in another way by defining the conformal factor as
\beq
\l = \exp \sigma :=
  \left( {\te \over {\Dp \r \Dm \r}} \right)^{1/2}
\eeq
Due to the $\r$-dependent modification, $\l$ transforms as
a scalar, i.e. a $(0,0)$ differential. Modulo fermionic
terms from \rf{box rho}, we have
\beq
\rD_{\dot \pm} \rD_{\dot \pm} \r =
- 2 {\cal D}_{\dot \pm} \sigma {\cal D}_{\dot \pm} \r
\labelx{Ds Dr}
\eeq
As already remarked in \cite{Schlad}, this result suggests an
interpretation of the fields $\r$ and $\sigma$ as longitudinal
target space degrees of freedom.

The above equations illustrate the ``back reaction" of matter
on the geometry. In contrast to conformal field theory,
where one has only the analog of the (super) Virasoro conditions
\rf{Virasoro} and \rf{2d RS3}, we now get the extra equations
\rf{2d curly Einstein} and \rf{2d RS1}, where the matter fields
act as ``sources" for the topological degrees of freedom.
It is not clear whether and how these equations
restrict the geometry. In string theory, the moduli and
supermoduli can be freely chosen and are integrated over only
after one has calculated the relevant string amplitudes in the
background provided by them. Here, they seem to partake in the
dynamics in a less trivial fashion. Although \rf{2d curly Einstein}
can be viewed merely as an equation determining the conformal
factor, it could conceivably restrict the (super)moduli
space associated with the inequivalent Lorentzian world sheets
\footnote{Perhaps the analogy with $4d$ black hole
resulting from the collapse of a massive star is useful here.
Whether or not this collapse takes place depends critically
on the initial mass (and velocity) distribution of the star.
Thus, the matter degrees of freedom affect the topology of the
ambient space-time at least via the initial conditions.}.
It is also not clear how to treat
the various equations of motion at the singular points of the
worldsheet, where $\cR (x) \propto \delta^{(2)} (x-x_0)$ ($x_0^\mu$
are the coordinates of the singular point). Setting $\mpm = \mmp =0$
for simplicity, we see that one way to satisfy
\rf{2d curly Einstein} is to require $P^A_{\dot \pm} \propto
\delta (x^{\dot \pm} - x_0^{\dot \pm} )$. Since $P^A_{\dot \pm} =
\pa_{\dot \pm} \ve^A + ...$, where $\ve^A$ are the basic
scalar fields and the dots stand for non-linear terms,
it follows that the scalar fields must have a
jump at the singular point\footnote{It is perhaps
no coincidence that in closed string field theory
a similar discontinuity occurs at the point where a string
splits in two. See e.g. \cite{GrossPer}.}.

The variations under local supersymmetry transformations with
parameters $\ep I$ and $\em I$ can be arrived at in a similar
fashion. For their derivation from \rf{3d supervariations}
a compensating $SO(1,2)$ rotation with parameter $\L_{2 \pm}=
-i \bar \e^I \g_\pm \p_2^I$ is necessary to maintain
the triangular form of the gauge \rf{dreibein}.
For the gravitino components, we deduce
\beql
\delta \smpm I = \rDp \em I   \;\;\; &,& \;\;\;
\delta \smmp I = \rDm \ep I  \zeile
\delta \thp I  =  \wurzelhalb \rDp \ep I \;\;\; &,& \;\;\;
\delta \thm I  =  \wurzelhalb \rDm \em I \zeile
\delta \pzp I  =  \ft1{\sqrt{2}} i \r^{-1} \Dp \r \, \ep I \;\;\;
           &,&            \;\;\;
\delta \pzm I  =  -  \ft1{\sqrt{2}} i \r^{-1} \Dm \r \, \em I
\eeql
while for the dreibein components, the result is
\beql
\delta \mpm &=& \wurzelzwei i \em I \smpm I (\MM ) \zeile
\delta \mmp &=& - \wurzelzwei i \ep I \smmp I (\MM ) \zeile
\itepp \delta \tepp &=& - 2i \ep I \thp I  \; + \;
      {1\over {\MM}} \, \mmp \delta \mpm  \zeile
\itemm \delta \temm &=& +2i \em I \thm I  \; + \;
      {1\over {\MM}} \, \mpm \delta \mmp   \zeile
\r^{-1} \delta \r &=& - \ep I \pzp I - \em I \pzm I
\eeql
In the matter sector, we find
\beq
\delta \chp \Adot = \ft1{\sqrt{2}} i \GI A \Adot \ep I
               \cP_\dop^A   \;\;\; , \;\;\;
\delta \chm \Adot = \ft1{\sqrt{2}} i \GI A \Adot \em I \cP_\dom^A
\eeq
and
\beq
{\V}^{-1} \delta \V = \left( - \ep I \chp \Adot + \em I \chm \Adot
                \right) \GI A \Adot  Y^A
\eeq

\Section{The Linear System}

We now generalize the linear system of \cite{NW}, employing
the conformal calculus developed in section 2.
As explained in \cite{BM, Schlad}, the construction of
the linear system requires the replacement of the matrix
$\V (x)$ by another matrix $\Vh$ depending on a spectral
parameter $t$, viz.
\beq
\V (x) \longrightarrow \Vh (x,t)
\eeq
The occurrence of a spectral parameter in linear systems (Lax pairs)
for non-linear equations is, of course, a well known phenomenon.
However, the linear system constructed here possesses some
rather unusual properties: not only does the spectral
parameter $t$ depend on the dilaton field $\r$ as in the purely
bosonic theories (see \cite{BM, Schlad}), but it now also depends
on the topological degrees of freedom via the Beltrami and
super Beltrami differentials, see \rf{Dpmt} below. This feature
is entirely due to the interaction of the (super)gravitational
degrees of freedom with the matter fields, and distinguishes
locally supersymmetric integrable systems from flat
space models with or without rigid supersymmetry.
Moreover, the spectral parameter $t$, in terms of which the
emergence of affine Kac Moody algebras in these models
can be directly understood, now
becomes a dynamical quantity of its own because the
equations determining it themselves obey an integrability
constraint that gives rise to one of the equations of motion.

The linear system can be parametrized as follows
\beql
\Vh^{-1} \Dp \Vh &=& \ft12 \widehat {\c Q}^{IJ}_\dop X^{IJ} +
                   \widehat {\c P}^A_\dop Y^A   \zeile
\Vh^{-1} \Dm \Vh &=&  \ft12 \widehat {\c Q}^{IJ}_\dom X^{IJ} +
                   \widehat {\c P}^A_\dom Y^A
\labelx{Vhpm}
\eeql
where the hatted quantities $\widehat \cQ$ and
and $\widehat \cP$ depend on $t$ in contrast to $\cQ$ and $\cP$,
which do not (see \rf{curly QP}). They are given by
\beql
\widehat {\c Q}^{IJ}_\dop &=& {\c Q}^{IJ}_\dop \, - \,  \wurzelzwei
  {t\over{(1+t)^2}} \, \left( i\GIJ \Adot \Bdot \chp \Adot
\chp \Bdot \, +\,  8 \pzp {[I} \thp {J]} \right) \zeile
&& \;\;\;\;\;\;\; -16\wurzelzwei i{{t^2}\over{(1+t)^4}} \, \pzp I \pzp J
\; + \;  8 {t\over{(1-t)^2}} \, \pzm {[I} \p_\dop^{J] \dom } \zeile
\widehat {\c Q}^{IJ}_\dom &=& {\c Q}^{IJ}_\dom \, + \, \wurzelzwei
{t\over{(1-t)^2}} \,  \left( -i\GIJ \Adot \Bdot
  \chm \Adot \chm \Bdot \, + \,  8 \pzm {[I} \thm {J]} \right) \zeile
&& \;\;\;\;\;\; + 16\wurzelzwei i {{t^2}\over{(1-t)^4}} \, \pzm I \pzm J
\; -\; 8{t\over{(1+t)^2}} \, \pzp {[I} \p_\dom^{J] \dop }  \zeile
\widehat {\c P}^A_\dop &=&
{{1-t}\over{1+t}}\,P_\dop^A\,+\, 2\wurzelzwei i{{t(1-t)}\over {(1+t)^3}}
   \,  \GI A \Adot \chp \Adot \pzp I  \, - \,
4 {{t}\over {1-t^2}} \, \GI A \Adot \chm \Adot \p_\dop^{I \dom} \zeile
\widehat {\c P}_\dom^A &=&
 {{1+t}\over{1-t}} \, P_\dom^A \, - \,  2\wurzelzwei i{{t(1+t)}\over {(1-t)^3}}
  \,   \GI A \Adot \chm \Adot \pzm I  \, - \,
  4{{t}\over {1-t^2}} \, \GI A \Adot \chp \Adot \p_\dom^{I \dop}
\labelx{LS}
\eeql
where $\lbrack I,J \rbrack $ denotes antisymmetrization in the indices
$I,J$ with strength one. A somewhat lengthy calculation now
establishes that, with the exceptions described below,
all equations of motion given in the
preceding section as well as the integrability
condition \rf{3d integrability} can be obtained by imposing the
generalized integrability constraint
\beq
\rDp \big( \Vh^{-1} \Dm \Vh \big) - \rDm \big( \Vh^{-1} \Dp \Vh \big)
+ \big[ \Vh^{-1} \Dp \Vh \, , \, \Vh^{-1} \Dm \Vh \big] = 0
\eeq
Note that the derivatives to the left are covariant, since
otherwise we would have to include a commutator term
$\Vh^{-1} \big[ \Dp , \Dm \big] \Vh $ on the right hand side.
In addition, one must make use of the
following set of differential equations for the spectral parameter
\beql
t^{-1} \Dp t &=& {{1-t}\over{1+t}} \, \r^{-1} \Dp \r  \; - \;
    {{4t}\over{1-t^2}} \p_\dop^{I \dom} \pzm I  \zeile
t^{-1} \Dm t &=& {{1+t}\over{1-t}} \, \r^{-1} \Dm \r  \; + \;
    {{4t}\over{1-t^2}} \p_\dom^{I \dop} \pzp I
\labelx{Dpmt}
\eeql
Since these are first order equations, their solution $t=t(x,w)$
involves one integration constant $w$.
We stress that the linear system \rf{LS}
gives rise to {\it all} fermionic field equations, whereas
the super Virasoro conditions \rf{2d RS3} were missed in
\cite{NW}. The only equations of motion that cannot be recovered
from \rf{LS} are \rf{box rho}, \rf{2d curly Einstein}, \rf{Virasoro}
and the Maxwell equation for $A_\mu$, i.e. precisely
the equations obtained by dimensional reduction of the
$3d$ Einstein equations \rf{3d symmetric Einstein}. Remarkably, however,
the equations \rf{Dpmt} are themselves subject to an
integrability constraint that yields one of the missing equations!
Namely, for
\beql
 \rDm (t^{-1} \Dp t ) - \rDp (t^{-1} \Dm t ) &=&  \zeile
 = - {{4t}\over{1-t^2}} \r^{-1} \Big( \rDp \rDm \r  &+&
\rDp \big( \r \smmp I \pzp I \big) +
\rDm \big( \r \smpm I \pzm I \big) \Big)
\eeql
to vanish we must impose \rf{box rho}. To recover the
equations of motion \rf{2d curly Einstein}
and \rf{Virasoro}, it has been proposed in \cite{BM} to incorporate
the conformal factor into the linear system replacing the
matrix $\Vh$ by the pair $(\l ,\Vh )$; due to the presence of
a central charge in the Kac Moody algebra \cite{Julia2},
the multiplication of two such pairs involves a non-trivial
group two-cocycle. However, this proposal has so far only been
shown to work for the bosonic theories in the special gauge
\rf{superconformal gauge}. We have so far not found a way to include
the Maxwell equation into the linear system \rf{LS}. Nonetheless, these
observations strongly suggest that there exists yet another
generalization of \rf{LS} that also gives rise to the remaining
equations of motion and that includes the spectral parameter as
one of the dynamical fields. The dependence of
$t$ on the topological degrees of freedom has not been
considered in earlier work where the relevant field configurations
were assumed to be asymptotically flat for the Euclidean
reduction and topologically trivial for colliding plane waves.
Observe also that the poles at $t=-1$ and $t=+1$ in \rf{LS}
and \rf{Dpmt} are associated with the positive and negative
chirality components of the bosonic and fermionic fields, respectively.

As in \cite{NW}, we can also reformulate local supersymmetry as
a Kac Moody type gauge transformation. Namely, defining
\beq
  \Vh^{-1} \delta \Vh  = \Vh^{-1} \delta_+ \Vh   \; + \;
                         \Vh^{-1} \delta_- \Vh
\eeq
with
\beql
\Vh^{-1} \delta_+ \Vh &:=&
    -8 {t\over{(1+t)^2}} \, \ep I \pzp J \; \halb X^{IJ} \; - \;
{{1-t}\over{1+t}} \, \GI A \Adot \ep I \chp \Adot  \, Y^A  \zeile
\Vh^{-1} \delta_- \Vh &:=&
    +8 {t\over{(1-t)^2}} \, \em I \pzm J \; \halb X^{IJ}  \; + \;
 {{1+t}\over{1-t}} \, \GI A \Adot \em I \chm \Adot \, Y^A
\labelx{KM supervariation}
\eeql
one can check that
\beql
\delta \left( \Vh^{-1} \Dp \Vh \right) = \Dp \left( \Vh^{-1} \delta
      \Vh \right)  +  \Big[ \Vh^{-1} \Dp \Vh \, ,\, \Vh^{-1} \delta
      \Vh \Big] - \zeile
    -   \delta \mpm \, \Vh^{-1} \pa_\dom \Vh  -
 8 {t\over{(1+t)^2}} \, \ep I \cS_\dop^J \, \ft12 X^{IJ} \zeile
\delta \left( \Vh^{-1} \Dm \Vh \right) = \Dm \left( \Vh^{-1} \delta
      \Vh \right)  +  \Big[ \Vh^{-1} \Dm \Vh \, ,\, \Vh^{-1} \delta
      \Vh \Big] - \zeile
      -    \delta \mmp \, \Vh^{-1} \pa_\dop \Vh     -
 8 {t\over{(1-t)^2}} \, \em I \cS_\dom^J \, \ft12 X^{IJ}
\labelx{delta VhdV}
\eeql
This means that, modulo the super Virasoro conditions \rf{2d RS3},
local supersymmetry transformations can be entirely encoded
into the Kac Moody gauge parameter \rf{KM supervariation}. In order
to obtain this result, the spectral parameter must also be varied
\beq
t^{-1} \delta t =
- {{1-t}\over{1+t}}\, \ep I \pzp I - {{1+t}\over {1-t}}\, \em I \pzm I
\eeq
This equation can either be proven by demanding \rf{delta VhdV}
to hold, or by checking its compatibility with \rf{Dpmt} and
the supersymmetry variations listed at the end of the
preceding section.

\Section{Outlook}
As explained in \cite{BM, Schlad}, the space of
stationary axisymmetric or colliding plane wave solutions
can be identified with the infinite dimensional coset speace
\beq
{\cal M}_{restr} = G^\infty / H^\infty
\eeq
where $G^\infty$ is the Kac Moody group corresponding to the
group $G$ (with $G=SL(2,{\bf R})$ for pure gravity and $G=E_8$
for $N=16$ supergravity) and depends on the constant spectral
parameter $w$, and $H^\infty$ is its ``maximal compact subgroup".
The precise definition of $H^\infty$ and the
coset space $\cM_{restr}$ is, however, somewhat
subtle due to the $x$-dependence of $t$.
E.g. for $G=SL(n,{\bf R})$ and $H=SO(n)$,
$H^\infty$ is defined to be the set of matrices
$h(x,t) \in G$, which is invariant under the Cartan type involution
\cite{Julia2, BM}
\beq
\tau^\infty: h(x,t) \longrightarrow h^T (x,{1\over t})
\eeq
{}From \rf{LS}, one can verify that the involution $\tau^\infty$
leaves the expressions $\Vh^{-1} {\cal D}_{\dot \pm} \Vh$ invariant,
which therefore belong to the Lie algebra of $H^\infty$.
The groups $G^\infty$ and $H^\infty$ act on $\Vh$ according to
\beq
\Vh (x,t) \longrightarrow g^{-1} (w) \Vh (x,t) h(x,t)
\eeq
generalizing the action (3.1) of the corresponding finite dimensional
groups $G$ and $H$ on ${\cal V} (x)$.
The elements of the coset space ${\cal M}_{restr}$ are then
defined to be the equivalence classes of matrices
$\Vh (x,t)$ with respect to the ``gauge group" $H^\infty$.
In view of the fact that $G^\infty$ ``does not know" about $x$,
it is quite remarkable how the $x$-dependence of the elements
of ${\cal M}_{restr}$, and thereby of the solutions of the
gravitational field equations, emerges from this definition.

To overcome the restriction to topologically trivial solutions
and to incorporate configurations involving the
topological degrees of freedom, a bigger coset space is obviously
needed. From string theory we know that the configuration space
of pure $2d$ gravity is nothing but the moduli space
$\cM_0$ of the corresponding Riemann surface (this is a finite
dimensional space at each genus, but since we are interested in
solutions for arbitrary genus, a universal moduli space of the
type discussed in \cite{FS} would perhaps be more appropriate).
Defining the total ``moduli space of solutions" as
\beq
\cM  := {{\rm solutions \; of \; field \; equations} \over
            {\rm gauge \; transformations}}
\eeq
we see that $\cM$ must contain both $\cM_0$ as well as
$\cM_{restr}$. Now, owing to the ``back reaction" of matter on
the geometry discussed previously, it seems very unlikely that
$\cM$ is the direct product of $\cM_0$ and $\cM_{restr}$.
A most intriguing question is whether $\cM$ can be represented
as a coset space like $\cM_{restr}$ above, but now with
bigger groups $G^{\infty \infty} \supset G^\infty$ and
$H^{\infty \infty} \supset H^\infty$. It appears likely, however,
that this question cannot be settled before yet another extension
of the linear system involving the Kaluza Klein vector $A_\mu$
and its equation of motion has been found.

In \cite{Schlad}, the conserved Kac Moody current was shown
to take the form
\beq
\cJ^\mu = \e^{\mu \nu} \pa_\nu \left( {{\pa \Vh}\over {\pa w}}
                     \Vh^{-1} \right)
\eeq
The associated conserved charges are given by
\beq
\int \left( \cJ_\dop \Dxp +  \cJ_\dom \Dxm    \right)
\eeq
where the integral is to be performed along a spacelike
``hyper-surface" $x^0 = const$. On a topologically
non-trivial Lorentzian world sheet, this set may decompose into
several disconnected components, and consequently there may be more
than one conserved charge at a given instant. The algebraic
structure and the interrelation between these charges remain to be
understood\footnote{I am grateful to K. Pohlmeyer for a discussion
on this point and for alerting me to \cite{Kornhass}, where this
phenomenon has been studied in a somewhat different context.}.

{\bf Acknowledgements:} I would like to thank L. Baulieu,
R.W. Gebert, O. Lechtenfeld, D. Maison, J. Teschner, and
especially R. Dick for numerous and enlightening discussions
in connection with this work.

\Appendix A {Some Useful Formulas}
For the dimensional reduction, we use the metric $\eta_{+-} =1,
\eta_{22} = -1$ together with $\e^{2+-} = \e_{2+-} = 1$.
Furthermore, we have the following representation
of the gamma matrices in two dimensions
\beq
\g_+ = \pmatrix{0&0 \cr \sqrt{2}  &0 \cr}   \;\;\; , \;\;\;
\g_- = \pmatrix{0& \sqrt{2} \cr 0 &0 \cr}   \;\;\; , \;\;\;
\g^3 = \pmatrix{1&0 \cr 0&-1 \cr}
\eeq
as well as $\g_2 = - \g^2 = i \g^3$. Thus,
\beq
\g^{+-} = \g_{-+} := \ft12 \big[ \g_- , \g_+ \big] = \g^3
\eeq
The charge conjugation matrix $\c C$ obeys $\c C^{-1} \g_\pm \c C =
- \g_\pm^T $ and $\c C^{-1} \g^3 \c C = - \g^3$.
We identify the real one-component spinors $\chi_\pm$ with the
components of the two-component spinor $\chi$, i.e. $\chi =
\pmatrix{ \chi_+ \cr \chi_- \cr}$. These are one-dimensional
representations of the local Lorentz group $SO(1,1)$, scaling
as $\chi_\pm \rightarrow e^{\pm {\a /2}} \chi_\pm$ under the action of
$SO(1,1)$ (if the Lorentz group were $SO(2)$, the one-component spinors
would scale with opposite complex phase factors $e^{\pm i{\a /2}}$
instead, hence would be complex). It is now straightforward
to check that
\beq
 \bar \ve \chi = \ve_+ \chi_- - \ve_- \chi_+ = \bar \chi \ve \;\;\; ,
 \;\;\; \bar \ve \g^3 \chi = -\ve_+ \chi_- - \ve_- \chi_+ =
       - \bar \chi \g^3 \ve
\eeq
and
\beq
\bar \ve \g_+ \chi = \wurzelzwei \ve_+ \chi_+  \;\;\; , \;\;\;
\bar \ve \g_- \chi = - \wurzelzwei \ve_- \chi_-
\eeq
where the components $\chi_\pm$ and $\e_\pm$ are treated as
anticommuting (i.e. Grassmann) variables in order for the required
symmetry properties under interchange to hold.

The coefficients of anholonomy are defined by
\beq
\O_{abc} := \euo a m \euo b n \big( \pa_m e_{nc}
                - \pa_n e_{mc} \big)
\eeq
and the spin connection is given by
\beq
\o_{abc} := \ft12 \big( \O_{abc}
           -\O_{bca} + \O_{cab} \big)
\eeq
in our conventions.

\end{document}